\documentclass[a4paper,11pt]{article}
\usepackage{pos}

\manuallySeparateAuthors

\newcommand*\circled[1]{\tikz[baseline=(char.base)]{
            \node[shape=circle,draw,inner sep=1pt] (char) {\tiny #1};}}

\usepackage{tikz}

\newcommand{\lsp}{\Lambda_{\cal S}}

\newcommand{\kappaOp}{Q}

\newcommand{\Mod}[1]{\ (\mathrm{mod}\ #1)}

\title{Fifty ways to build a deuteron: a variational calculation of two-nucleon systems}
\ShortTitle{Fifty ways to build a deuteron}

\author*{Michael~L.~Wagman}
\author[1]{ for the NPLQCD Collaboration}
\note{The research presented here was performed in collaboration with Saman Amarasinghe, Riyadh Baghdadi, Zohreh Davoudi, William Detmold, Marc Illa, Assumpta Parre\~no, Andrew Pochinsky, and Phiala Shanahan.}

\affiliation{Fermi National Accelerator Laboratory, \\ Batavia, IL 60510, USA}

\emailAdd{mwagman@fnal.gov}
\abstract{A variational study of two-nucleon systems with lattice quantum chromodynamics is performed using a wide range of interpolating operators: dibaryon operators built from products of momentum-projected nucleons, hexaquark operators built from six spatially localized quarks, and quasi-local operators inspired by two-nucleon bound-state wavefunctions in nuclear effective field theories. Correlation-function matrices involving products of these operators are constructed by computing timeslice-to-all quark propagators with sparsening techniques. Comparisons between results obtained using the same gauge-field ensemble but different interpolating-operator sets demonstrate that interpolating-operator dependence can lead to significant effects on the two-nucleon energy spectra obtained using both variational and non-variational methods. }

\FullConference{%
 The 38th International Symposium on Lattice Field Theory, LATTICE2021
  26th-30th July, 2021
  Zoom/Gather@Massachusetts Institute of Technology
}

\begin{document}
\maketitle

\section{Introduction}

Accurately disentangling ground- and excited-state effects in Euclidean correlation functions is a critical but challenging step in lattice quantum chromodynamics (LQCD) calculations of the energy spectra of hadrons and nuclei, matrix elements of electroweak and beyond-Standard-Model currents, and other observables.
For asymptotically large Euclidean time separations $t$, correlation functions involving arbitrary ``source'' and ``sink'' operators approach proportionality to $e^{-t E_0}$ with excited-state effects suppressed by $O( e^{-t \delta} )$, where $\delta \equiv E_1 - E_0$ is the energy gap between the ground state and the first excited state. % and $z_{\mathsf{n}}$ is the ratio of overlap factors between this excited state and the ground state for a given interpolating operator.
Although large Euclidean time separations can therefore be used to suppress excited-state effects in principle, LQCD calculations of multi-nucleon systems are particularly computationally challenging because of an exponential signal-to-noise problem that becomes more severe for larger $t$, lighter quark masses, and larger baryon-number systems~\cite{Parisi:1983ae, Lepage:1989hd, Beane:2009gs, Wagman:2016bam, Davoudi:2020ngi}.
Energy gaps between the ground state and excited states are small for nuclear systems in large volumes, because unbound finite volume (FV) states have energies that approach threshold like powers of $1/L$ for large spatial extents $L$~\cite{Luscher:1986pf}.
The combination of these effects has made achieving LQCD calculations with $t \gg \delta^{-1} $ unfeasible with current algorithms.\footnote{An alternative approach based on determining nuclear potentials from Bethe-Salpeter wavefunctions of multi-baryon systems~\cite{Ishii:2006ec, Inoue:2010es} is argued to avoid the need to achieve $t \gg \delta^{-1}$ to remove contamination from ``elastic'' excited states~\cite{Aoki:2020bew}; however, short-distance features of the resulting potentials depend on the choice of sink interpolating operator~\cite{Birse:2012ph} and the systematic uncertainties in this approach cannot be quantitatively estimated~\cite{Beane:2010em, Birse:2012ph, Yamazaki:2017gjl, Iritani:2018zbt, Drischler:2019xuo}.}
%estimated~\cite{Beane:2010em, Birse:2012ph, Walker-Loud:2014iea, Kawai:2017goq, Yamazaki:2017gjl, Davoudi:2017ddj, Iritani:2018zbt, }.
%including a dependence on the choice of interpolating operator used to annihilate the two-nucleon state appearing in the Bethe-Salpeter wavefunctions that allows inconsistent sets of potentials to be obtained when the approach is applied to different interpolating operators in toy models~\cite{Birse:2012ph}.
Understanding the systematic uncertainties associated with determinations of FV energy spectra determined using correlation functions with $t \lesssim \delta^{-1}$ is crucial for providing LQCD calculations of nuclear systems with fully quantified uncertainties.

Variational methods, in which solutions to generalized eigenvalue problems (GEVP) are used to obtain orthogonal sets of approximate energy eigenstates, provide determinations of energy spectra from correlation functions that are guaranteed to provide upper bounds on the true energy spectra for arbitrary $t$~\cite{Fox:1981xz,Michael:1982gb,Luscher:1990ck}.
%Further, orthogonality of GEVP eigenstates guarantees that ground-state energy determinations using variational methods are free from contamination from the lowest-energy set of states that significantly overlap with the interpolating operators used in a particular calculation. 
If a set of interpolating operators that significantly overlaps with all energy eigenstates below a threshold excitation energy $\Delta$ can be constructed, then variational methods can suppress excited-state effects on ground-state energy determinations from $O(e^{-t \delta})$ to $O(e^{-t \Delta})$~\cite{Luscher:1990ck}.
However, since the structure of QCD energy eigenstates is not known \emph{a priori}, it is non-trivial to determine whether a given interpolating operator set has sufficient overlap with all low-energy states below a given $\Delta$ and thereby quantitatively estimate the excited-state contamination present in ground-state energy determinations.

The first variational LQCD studies investigated the baryon-number $B=0$ and $B=1$ sectors using single-hadron glueball, meson, and baryon interpolating operators~\cite{Michael:1982gb,Allton:1993wc,Burch:2004he}.
Construction of multi-hadron interpolating operators is more computationally challenging and was enabled more recently through the development of efficient algorithms for calculating approximate all-to-all quark propagators, namely the Laplacian Heaviside or ``distillation'' method~\cite{Peardon:2009gh} and the stochastic Laplacian Heaviside~\cite{Morningstar:2011ka} method.
It was observed for both $B=0$~\cite{Dudek:2012xn,Wilson:2015dqa} and $B=1$~\cite{Lang:2012db,Kiratidis:2015vpa} systems with energies close to resonances or particle production thresholds that interpolating-operator sets including only single-hadron or multi-hadron operators can lead to inconsistent determinations of energy spectra with ``missing energy levels'' when compared to results obtained using larger interpolating-operator sets.

%TThe application of variational methods to $B \geq 2$ systems is required to enable similarly detailed studies of multi-nucleon energy spectra.
Early calculations of the energy spectra of $B \geq 2$ systems~\cite{Beane:2010hg,Beane:2011iw,Beane:2012vq,Yamazaki:2012hi,Beane:2013br,Berkowitz:2015eaa,Yamazaki:2015asa,Wagman:2017tmp} and nuclear matrix elements reviewed in Ref.~\cite{Davoudi:2020ngi}
%,Beane:2015yha,Chang:2015qxa,Detmold:2015daa,Parreno:2016fwu,Savage:2016kon,Shanahan:2017bgi,Tiburzi:2017iux,Winter:2017bfs,Chang:2017eiq,Detmold:2020snb} 
employed asymmetric correlation functions involving single-hadron sources and multi-hadron sinks that can be calculated using efficient contraction algorithms~\cite{Doi:2012xd,Detmold:2012eu} but do not provide variational upper bounds on energy levels.
The first variational study of $B=2$ systems was reported by Francis et al. in Ref.~\cite{Francis:2018qch} and performed calculations in several boosted frames of $2\times 2$ positive-definite, Hermitian correlation-function matrices involving single-hadron interpolating operators as well as symmetric multi-hadron correlation functions computed using the stochastic Laplacian Heaviside method.
Variational studies have also been performed by H{\"o}rz et al. in Ref.~\cite{Horz:2020zvv} and Green et al. in Ref.~\cite{Green:2021qol} in which 2 or 3 multi-hadron interpolating operators were used to build positive-definite Hermitian correlation-function matrices in several boosted frames.
The results of these variational studies are in tension with results from asymmetric correlation functions, although different calculations involve different discretizations and quark masses, and it was suggested in Ref.~\cite{Green:2021qol} that lattice spacing artifacts may contribute to the discrepancies.

This work presents a variational study of $B=2$ systems in which sparsening methods~\cite{Detmold:2019fbk,Li:2020hbj} and highly optimized codes using the \verb!Tiramisu!~\cite{baghdadi2020tiramisu} compiler framework are used to enable efficient calculations of positive-definite Hermitian correlation-function matrices
%Calculations are restricted to positive parity systems with center-of-mass momentum $\vec{P}=\vec{0}$, and correlation-function matrices are decomposed  into blocks with definite total isospin, $I$, that transform in irreducible representations (irreps), $\Gamma_J$, of the (double cover of the) cubic group that play the role of total angular momentum for these FV systems.
 including both single- and multi-hadron interpolating operators with dimensionalities as large as $16\times 16$ in the isospin $I=1$ ``dineutron'' channel and $42\times 42$ in the $I=0$ ``deuteron'' channel.
The asymmetric correlation functions studied in previous works appear as particular off-diagonal entires of these correlation-function matrices and a subset of the gauge-field ensemble used in Refs.~\cite{Beane:2012vq,Berkowitz:2015eaa,Wagman:2017tmp} is used here, enabling direct comparisons of results obtained using asymmetric correlation functions and variational methods on the same gauge-field ensemble.
The methods and results of this study are summarized below and detailed in Ref.~\cite{Amarasinghe:2021lqa}.

\section{Interpolating operators}

\subsection{Nucleon operators}

Standard proton interpolating operators with $B=1$ and $I=1/2$ that transform in the $\Gamma_J = G_1^+$ irrep of the double-cover of the cubic group can be expressed in the Dirac basis (see Ref.~\cite{Basak:2005ir}) as
\begin{equation}
\begin{split}
  p_{\sigma g}(x) &= \varepsilon^{abc} \frac{1}{\sqrt{2}}\left[ u_{\zeta g}^{a}(x) (C \gamma_5 P_+)_{\zeta \xi} d_{\xi g}^{b}(x) - d_{\zeta g}^{a}(x) (C \gamma_5 P_+)_{\zeta \xi} u_{\xi g}^{b}(x) \right] \\
   &\hspace{20pt} \times \left[ P_+ \left( 1 - (-1)^\sigma i \gamma_1 \gamma_2 \right) \right]_{\sigma \zeta} u_{\zeta g}^{c}(x),
  \end{split}%\label{eq:Ninterp}
    \label{eq:Nsmear_def}
\end{equation}
where $\sigma \in \{0,1\}$ labels the row of $G_1^+$ corresponding to the proton spin, $q_{\zeta g}^a (x)$ denotes a quark field of flavor $q \in \{u,d\}$ with  $SU(3)$ color indices $a,b,c$,  Dirac spinor indices $\zeta, \xi$, and labels $g$ specifying the gauge-invariant Gaussian smearing~\cite{Gusken:1989ad},  % $\gamma_\mu$ are Euclidean gamma matrices, %satisfying $\gamma_\mu^\dagger =\gamma_\mu$ and $\{\gamma_\mu,\gamma_\nu\}=2\delta_{\mu\nu}$, 
$C=\gamma_2\gamma_4$, $\gamma_5=\gamma_1 \gamma_2 \gamma_3 \gamma_4$, and $P_+ \equiv \left( \frac{1+\gamma_4}{2} \right)$ is a positive-parity projector.
Neutron operators $n_{\sigma g}(x)$ are defined by Eq.~\eqref{eq:Nsmear_def} with $u\leftrightarrow d$, and the isodoublet nucleon operator is $N_{\sigma g} \equiv (p_{\sigma g}(x), n_{\sigma g}(x))^T$.

Projection to a definite center-of-mass momentum $\vec{P}_{\mathfrak{c}}$ is accomplished by multiplying $N_{\sigma g}(\vec{x},t)$ by $e^{i \vec{P}_{\mathfrak{c}} \cdot \vec{x}}$ and summing over the set of spatial lattice sites $\Lambda$.
The propagator sparsening algorithm introduced in Ref.~\cite{Detmold:2019fbk} significantly reduces the computational cost of this summation and more costly summations arising for multi-hadron interpolating operators.
Sparsened plane-wave spatial wavefunctions are defined to only have support on a cubic sublattice $\Lambda_{\mathcal{S}} \subset \Lambda$ defined as
\begin{equation}
    \lsp = \{(x_1,x_2,x_3)\ |\ 0 \leq x_k < L,\ x_k \equiv 0 \Mod{\mathcal{S}}\},
\label{eq:sparse_lattice}
\end{equation} 
where lattice units are used and $\mathcal{S} \in \mathbb{Z}$ is the ratio of the number of full and sparse lattice sites in each spatial dimension.
Momentum-projected nucleon operators are defined as
\begin{equation}
  N_{\sigma\mathfrak{c}g}(t) = \sum_{\vec{x}\in \lsp} \psi_\mathfrak{c}^{[N]}(\vec{x}) N_{\sigma g}(\vec{x},t), \label{eq:Nprojdef}
\end{equation}
where $\psi_\mathfrak{c}^{[N]}(\vec{x}) \equiv \left. e^{i \vec{P}_{\mathfrak{c}}} \right|_{\Lambda_{\mathcal{S}}}$ is the spatial wavefunction with support restricted to the sparse lattice.
Sparsening with $\mathcal{S} > 1$ leads to incomplete Fourier projection in which operators overlap with momenta whose components differ from $\vec{P}_{\mathfrak{c}}$ by multiples of $\frac{2\pi}{\mathcal{S}}$.
%$\vec{P}_{\mathfrak{c}} \pm \left( \frac{2\pi}{\mathcal{S}} \right) \hat{e}_k$ where $\hat{e}_k$ is a spatial unit vector.
Sparsening therefore leads to the appearance of additional excited-state contamination in correlation functions, but these effects can be mitigated by taking $t \gg \frac{2\pi}{\mathcal{S}}$, which,  assuming $\mathcal{S} \ll L$, is easier to achieve in practice than suppression of excited-state effects in multi-hadron correlation functions.
%For example for a nucleon with $\vec{P}_\mathfrak{c} = \vec{0}$ these effects are suppressed by $e^{-t \Delta E_{\mathcal{S}}^{(1,\frac{1}{2},G_1^+)} }$, where $\Delta E_{\mathcal{S}}^{(1,\frac{1}{2},G_1^+)} =\sqrt{ M_N^2 + \left( \frac{2\pi}{\mathcal{S}} \right)^2 } - M_N$, with $M_N \equiv E_0^{(1,\frac{1}{2},G_1^+)}$ the mass of the nucleon.

\subsection{Hexaquark operators}

Local six-quark operators, or ``hexaquark operators,'' can be constructed from products of two nucleon operators with Gaussian smeared quark fields centered about the same lattice site,
\begin{equation}
\begin{split}
  H_{\rho\mathfrak{c}g}(t) &= \sum_{\vec{x} \in \lsp} \psi_\mathfrak{c}^{[H]}(\vec{x}) \sum_{\sigma,\sigma'} v^\rho_{\sigma \sigma^\prime}\, \frac{1}{\sqrt{2}}\left[ p_{\sigma g}(\vec{x}, t) n_{\sigma' g}(\vec{x}, t) \right. \\
    &\hspace{180pt} \left. + (-1)^{1-\delta_{\rho 0}} n_{\sigma g}(\vec{x}, t) p_{\sigma' g}(\vec{x}, t)\right], 
\end{split}\label{eq:Hinterp}
\end{equation}
where $\psi_\mathfrak{c}^{[H]}(\vec{x}) \equiv \left. e^{i \vec{P}_{\mathfrak{c}}} \right|_{\Lambda_{\mathcal{S}}}$, $\rho \in \{0,\ldots,3\}$ labels the row of $G_1^+ \otimes G_1^+ = A_1^+ \oplus T_1^+$, and projection to spin-singlet ($\rho=0$) and spin-triplet ($\rho \in \{1,2,3\}$) operators is achieved using
\begin{equation}
\begin{split}
  v^{0}_{\sigma\sigma^\prime} &= \frac{1}{\sqrt{2}}(\delta_{\sigma 0} \delta_{\sigma^\prime 1} - \delta_{\sigma 1} \delta_{\sigma^\prime 0}), \hspace{20pt}
    v^1_{\sigma\sigma^\prime} = \delta_{\sigma 0} \delta_{\sigma^\prime 0}, \\
    v^{2}_{\sigma\sigma^\prime} &= \frac{1}{\sqrt{2}}(\delta_{\sigma 0} \delta_{\sigma^\prime 1} + \delta_{\sigma 1} \delta_{\sigma^\prime 0}), \hspace{20pt}
    v^3_{\sigma\sigma^\prime} = \delta_{\sigma 1} \delta_{\sigma^\prime 1}.
\end{split}\label{eq:veights}
\end{equation}
Hexaquark correlation functions are efficiently calculated by using quark-exchange symmetry to reduce the number of terms in a sparse tensor representation of the operators as in Ref.~\cite{Detmold:2012eu}.

\subsection{Dibaryon operators}

Six-quark operators can also be constructed from products of momentum projected baryon operators.
These ``dibaryon operators'' are defined analogously to Eq.~\eqref{eq:Hinterp} as
\begin{equation}
\begin{split}
  D_{\rho\mathfrak{m}g}(t) &= \sum_{\vec{x}_1,\vec{x}_2 \in \lsp} \psi_\mathfrak{m}^{[D]}(\vec{x}_1,\vec{x}_2) \sum_{\sigma,\sigma'} v^\rho_{\sigma \sigma^\prime}\, \frac{1}{\sqrt{2}}\left[ p_{\sigma g}(\vec{x}_1, t) n_{\sigma' g}(\vec{x}_2, t) \right. \\
    &\hspace{180pt} \left. + (-1)^{1-\delta_{\rho 0}} n_{\sigma g}(\vec{x}_1, t) p_{\sigma' g}(\vec{x}_2, t)\right], 
\end{split}\label{eq:Dinterp}
\end{equation}
which include bilocal spatial wavefunctions labeled by $\mathfrak{m}$ that are defined by
\begin{equation}
    \psi_\mathfrak{m}^{[D]}(\vec{x}_1,\vec{x}_2) = \frac{1}{\sqrt{2}} \left[ e^{i\left( \frac{\vec{P}_{\mathfrak{m}}}{2} + \vec{k}_{\mathfrak{m}}\right) \cdot \vec{x}_1 } e^{i\left( \frac{\vec{P}_{\mathfrak{m}}}{2} - \vec{k}_{\mathfrak{m}}\right) \cdot \vec{x}_2 } + e^{i\left( \frac{\vec{P}_{\mathfrak{m}}}{2} + \vec{k}_{\mathfrak{m}}\right) \cdot \vec{x}_2 } e^{i\left( \frac{\vec{P}_{\mathfrak{m}}}{2} - \vec{k}_{\mathfrak{m}}\right) \cdot \vec{x}_1 } \right].
\label{eq:psiDdef}
\end{equation}

To project dibaryon operators into cubic irreps, it is convenient to first form linear combinations of dibaryon operators within the same relative momentum ``shell'' $\mathfrak{s}(\mathfrak{m}) \equiv \vec{k}_{\mathfrak{m}} \cdot \vec{k}_{\mathfrak{m}}$ that lead to spatial wavefunctions that transform in a particular irrep $\Gamma_\ell$ (analogous to infinite-volume orbital angular momentum).
The linear combinations required for $\mathfrak{s} \leq 6$ are presented in Refs.~\cite{Luu:2011ep,Amarasinghe:2021lqa}.
Dibaryon operators with definite $\Gamma_J$, the FV analog of total angular momentum, can then be constructed using Clebsch-Gordon coefficients for $\Gamma_J = \Gamma_\ell \otimes \Gamma_S$, where $\Gamma_S$ is the irrep associated with the two-nucleon spin as for hexaquark operators above, which are presented in Ref.~\cite{Basak:2005ir}.
Sparsening leads to excited-state contamination from systems with center-of-mass momentum equal to $\vec{k}_{\mathfrak{m}} \pm \left(\frac{2\pi}{\mathcal{S}}\right)\vec{e}_i$ for $i\in\{1,2,3\}$. These effects are less suppressed than in the single-nucleon case, but they are still expected to be negligible compared to other excited-state effects for $\mathcal{S} \ll L$ as detailed in Ref.~\cite{Amarasinghe:2021lqa}.

It is computationally efficient to first calculate correlation-function matrices involving dibaryon operators with plane-wave spatial wavefunctions as in Eq.~\eqref{eq:Dinterp} and subsequently perform a change-of-basis that projects the source and sink operators to definite $\Gamma_J$.
This is because the wavefunctions in Eq.~\eqref{eq:psiDdef} factorize into (a sum of two) products of independent wavefunctions for each baryon, and correlation functions can therefore be calculated by first constructing ``baryon blocks'' in which the sums over the spatial locations of each baryon are performed independently.
Expressing correlation functions as a product of the resulting sums instead of a sum of products reduces the computational cost of constructing dibaryon correlation functions from $O(L_\mathcal{S}^{12})$ to $O(L_\mathcal{S}^9)$ and gives a factor of $10^5$ cost reduction in the numerical study described below.

\subsection{Quasi-local operators}

In low-energy effective thoeries and phenomenological nuclear models  with nucleon degrees of freedom, the deuteron is described as a loosely bound two-nucleon system with a spatial wavefunction that decays exponentially for large separations of the nucleons.
One may worry that both local hexaquark operators and dibaryon operators whose wavefunctions are not suppressed for large nucleon separations  might have small overlap with such a loosely bound state.
``Quasi-local'' interpolating operators that more closely resemble FV EFT expectations for the deuteron wavefunction can be defined by
\begin{equation}
\begin{split}
    \kappaOp_{\rho\mathfrak{q}g}(t) &= \sum_{\vec{x}_1,\vec{x}_2 \in \lsp} \psi_{\mathfrak{q}}^{[\kappaOp]}(\vec{x}_1,\vec{x}_2,\vec{R}) \sum_{\sigma,\sigma'} v^\rho_{\sigma \sigma^\prime}\, \frac{1}{\sqrt{2}}\left[ p_{\sigma g}(\vec{x}_1, t) n_{\sigma' g}(\vec{x}_2, t) \right. \\
    &\hspace{180pt} \left. + (-1)^{1-\delta_{\rho 0}} n_{\sigma g}(\vec{x}_1, t) p_{\sigma' g}(\vec{x}_2, t)\right], 
\end{split}\label{eq:Einterp}
\end{equation}
with wavefunctions with an exponential localization scale labeled by $\mathfrak{q}$
\begin{equation}
  \psi_{\mathfrak{q}}^{[\kappaOp]}(\vec{x}_1,\vec{x}_2,\vec{R})  = \frac{1}{L_{\mathcal{S}}^3} \sum_{\tau \in \mathbb{T}_{\mathcal{S}}} e^{ - \kappa_{\mathfrak{q}} |\tau(\vec{x}_1) - \vec{R}|}e^{ - \kappa_{\mathfrak{q}} |\tau(\vec{x}_2) - \vec{R}|},
  \label{eq:Edef}
\end{equation}
where $\mathbb{T}_{\mathcal{S}}$ is the set of translations by multiples of the sparse lattice spacing and $\vec{R}$ is an arbitrary parameter specifying the center of the two-nucleon system before translation averaging.
By performing the sum over translations only at the sink, correlation functions involving quasi-local operators can be constructed using wavefunctions that factorize into a product of independent spatial wavefunctions for each nucleon at the source and can therefore be efficiently calculated using the same baryon block algorithm that is applied to construct  dibaryon correlation functions.

\section{Variational methods}

Given a set $\mathbb{S}$ of interpolating operators generically denoted by $\chi, \chi'$, an equal-sized set of approximately orthogonal interpolating operators that dominantly overlap with a single energy eigenstate can be obtained by solving the GEVP~\cite{Michael:1982gb,Luscher:1990ck},
\begin{equation}
    \sum_{\chi'} C_{\chi\chi'}^{(B,I,\Gamma_J)}(t) v^{(B,I,\Gamma_J,\mathbb{S})}_{{\mathsf{n}}\chi'}(t,t_0)  = \lambda_{\mathsf{n}}^{(B,I,\Gamma_J,\mathbb{S})}(t,t_0) \sum_{\chi'} C_{\chi\chi'}^{(B,I,\Gamma_J)}(t_0) v^{(B,I,\Gamma_J,\mathbb{S})}_{{\mathsf{n}}\chi'}(t,t_0),
\label{eq:GEVP}
\end{equation}
where $(B,I,\Gamma_J)$ denotes the baryon number, isospin, and cubic irrep of each state,
$\lambda_{\mathsf{n}}^{(B,I,\Gamma_J,\mathbb{S})}(t,t_0)$ are the eigenvalues, $v^{(B,I,\Gamma_J,\mathbb{S})}_{{\mathsf{n}}\chi'}(t,t_0)$ are the eigenvectors, and $t_0$ is a reference time that can be for example a fixed $t$-independent value or a fixed fraction of $t$.
Correlation functions for these approximate energy eigenstates can then be constructed as
\begin{equation}
    \widehat{C}_{\mathsf{n}}^{(B,I,\Gamma_J,\mathbb{S})}(t) = \sum_{\chi\chi'} v_{{\mathsf{n}}\chi}^{(B,I,\Gamma_J,\mathbb{S})}(t_{\rm ref},t_0)^*  C_{\chi\chi'}^{(B,I,\Gamma_J)}(t) v^{(B,I,\Gamma_J,\mathbb{S})}_{{\mathsf{n}}\chi'}(t_{\rm ref},t_0),
\label{eq:GEVPcorrelators}
\end{equation}
where $t_{\rm ref}$ is a reference time. Taking $t_0$ and $t_{\rm ref}$ to be fixed parameters independent of $t$ guarantees that $\widehat{C}_{\mathsf{n}}^{(B,I,\Gamma_J,\mathbb{S})}(t)$ has a spectral representation with positive-definite contributions from each state,
\begin{equation}
\begin{split}
    \widehat{C}_{\mathsf{n}}^{(B,I,\Gamma_J,\mathbb{S})}(t) &= \sum_{\mathsf{m}}  \left|\sum_{\chi} v_{{\mathsf{n}}\chi}^{(B,I,\Gamma_J,\mathbb{S})}(t_{\rm ref},t_0)^* Z_{\mathsf{m}\chi}^{(B,I,\Gamma_J)}\right|^2 e^{-tE_{\mathsf{m}}^{(B,I,\Gamma_J)} },
\end{split}\label{eq:GEVPspectral}
\end{equation}
which guarantees that effective energies $E_{\mathsf{n}}^{(B,I,\Gamma_J,\mathbb{S})}(t) =  \ln \left( \frac{ \widehat{C}_{\mathsf{n}}^{(B,I,\Gamma_J,\mathbb{S})}(t) }{ \widehat{C}_{\mathsf{n}}^{(B,I,\Gamma_J,\mathbb{S})}(t+1) } \right)$ approach the true energies $E_{\mathsf{n}}^{(B,I,\Gamma_J)}$ from above with no possibility of ``false plateaus'' arising from cancellations between different excited states. Fits of $\widehat{C}_{\mathsf{n}}^{(B,I,\Gamma_J,\mathbb{S})}(t)$ to sums of exponentials can be used to provide energy-level determinations $E_{\mathsf{n}}^{(B,I,\Gamma_J,\mathbb{S})}$. FV energy shifts that can be used to constrain baryon-baryon scattering amplitudes can then be determined from linear combinations $\Delta E_{\mathsf{n}}^{(2,I,\Gamma_J,\mathbb{S})} = E_{\mathsf{n}}^{(2,I,\Gamma_J,\mathbb{S})} - 2 E_0^{(1,\frac{1}{2},G_1^+,\mathbb{S}')}$. Linear combinations of effective energies are defined analogously.
The relative overlaps of each interpolating operator onto each energy eigenstate, denoted $\mathcal{Z}_{\mathsf{n}\chi}^{(B,I,\Gamma_J,\mathbb{S})}$, are obtained using $E_{\mathsf{n}}^{(B,I,\Gamma_J,\mathbb{S})}$, $\widehat{C}_{\mathsf{n}}^{(B,I,\Gamma_J,\mathbb{S})}(t_{\rm ref})$, and $v_{{\mathsf{n}}\chi}^{(B,I,\Gamma_J,\mathbb{S})}(t_{\rm ref},t_0)$ as in Ref.~\cite{Bulava:2016mks}.

\subsection{Missing energy levels}\label{sec:toy}

For $t \rightarrow \infty$, any interpolating operator set will overlap with an equal-sized set of the lowest energy eigenstates and excited-state effects are exponentially suppressed by the energy gap to the lowest-energy state outside the space spanned by the interpolating operator set.
For finite $t$, variational methods are guaranteed to provide upper bounds on the true energy spectrum but will not provide accurate determinations of energy levels for states with small energy gaps (compared to $1/t$) that have sufficiently small overlap with all interpolating operators considered.
For example, consider a pair of interpolating operators $A$ and $B$ and a three-state system with true energy levels
\begin{equation}
  E_0^{(AB)} = \eta - \Delta, \hspace{20pt} E_1^{(AB)} = \eta, \hspace{20pt} E_2^{(AB)} = \eta + \delta,
\end{equation}
and normalized overlap factors for operators $A$ and $B$ onto these states
\begin{equation}
  \mathcal{Z}_A = (\epsilon,\sqrt{1 - \epsilon^2},0), \hspace{20pt} \mathcal{Z}_B = (\epsilon,0,\sqrt{1 - \epsilon^2}),
  \label{eq:badZ}
\end{equation}
with $\epsilon\ll 1$ a real parameter.
The eigenvalues obtained by solving the GEVP for the $2\times 2$ correlation-function matrix involving $A$ and $B$ are given to $O(\epsilon^3)$ accuracy by
\begin{equation}
  \begin{split}
  \lambda_0^{(AB)} &= e^{-(t-t_0)\eta}\left[ 1 + \epsilon^2  \left(  e^{t \Delta} - e^{t_0 \Delta} \right) + \mathcal{O}(\epsilon^4) \right], \\
    \lambda_1^{(AB)} &= e^{-(t-t_0)(\eta + \delta) }\left[ 1 + \epsilon^2 \left(  e^{t (\Delta + \delta)} - e^{t_0 (\Delta + \delta)} \right) + \mathcal{O}(\epsilon^4) \right].
  \end{split}
\end{equation}
The GEVP eigenvalues will therefore not recover the true ground-state energy unless $t$ is large enough that $e^{t \Delta}$ compensates for the $O(\epsilon^2)$ overlap-factor suppression.
An asymmetric correlation function involving $A$ and $B$ operators, however, will overlap perfectly with the true ground state with zero excited-state contamination.
This example can be trivially generalized to include more states and interpolating operators. % and provides a quantitative relation between the overlap of an interpolating operator set with a particular state and the $t$ required to ensure that the energy of this state can be reliably determined using variational methods.

\section{Numerical study}

A variational study including the operators above is performed using $N_{\rm cfg} = 167$ gauge-field configurations with $N_f =3$ degenerate quarks with $m_\pi = 806$ MeV, $L=32$, and $a = 0.1453(16)$ fm~\cite{Beane:2012vq}, corresponding to a subset of the configurations used in Refs.~\cite{Beane:2012vq,Berkowitz:2015eaa,Wagman:2017tmp}. The tadpole-improved~\cite{Lepage:1992xa} L{\"u}scher-Weisz gauge-field action~\cite{Luscher:1984xn} and the Wilson quark action including the Sheikholeslami-Wohlert (clover) improvement term~\cite{Sheikholeslami:1985ij} are used with one step of four-dimensional stout smearing~\cite{Morningstar:2003gk} with $\rho=0.125$ applied to the gauge field.
Sparsened timeslice-to-all quark propagators with $\mathcal{S} = 4$, corresponding to $L_{\mathcal{S}} = 8$, are computed using Gaussian smeared sources and sinks with two different widths denoted  below as $W$ for ``wide'' and $T$ for ``thin.'' 
The \verb!Qlua! LQCD software framework~\cite{qlua} is used to perform these calculations. 
Sparsened timeslice-to-all propagators are stored and subsequently used to calculate correlation-function matrices using \verb!C++! codes including significant scheduling and memory optimizations facilitated by the polyhedral compiler \verb!Tiramisu!~\cite{baghdadi2020tiramisu}.

For the nucleon, a $2\times 2$ correlation-function matrix including thin- and wide-smeared interpolating operators is used to construct GEVP correlation functions.
A similar fitting strategy is employed as in Ref.~\cite{Beane:2020ycc}: single- and multi-exponential fits with a variety of minimum $t$ included in fits are performed, an information criterion is used to determine the optimal number of states to include in each fit, and a weighted average of acceptable fit results is used to determine $E_{\mathsf{n}}^{(B,I,\Gamma_J,\mathbb{S})}$. 
The result for the nucleon mass, $E_0^{(1,\frac{1}{2},G_1^+,\mathbb{S}_N)} = 1.2031(30)$, is consistent with previous higher-precision determinations using the same gauge-field ensemble, which obtained $M_N = 1.20396(83)$~\cite{Wagman:2017tmp}.

\begin{figure}[!t]
	\includegraphics[width=0.49\columnwidth]{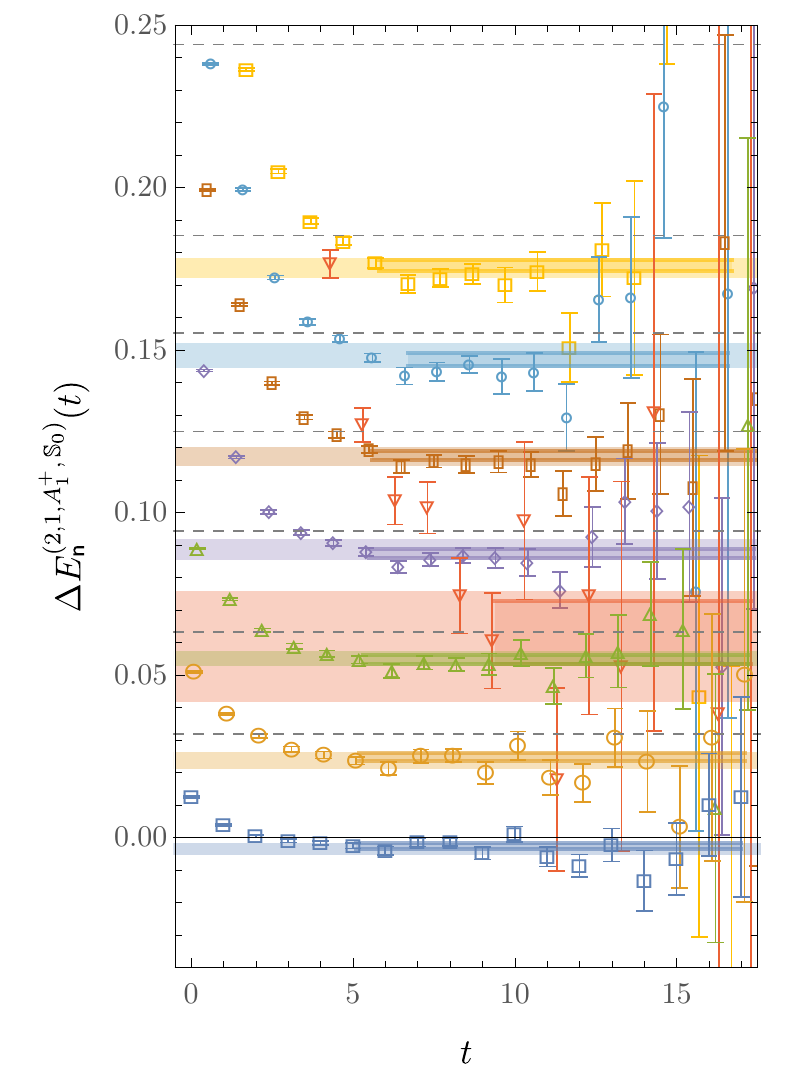} \hspace{4mm}
	\includegraphics[width=0.49\columnwidth]{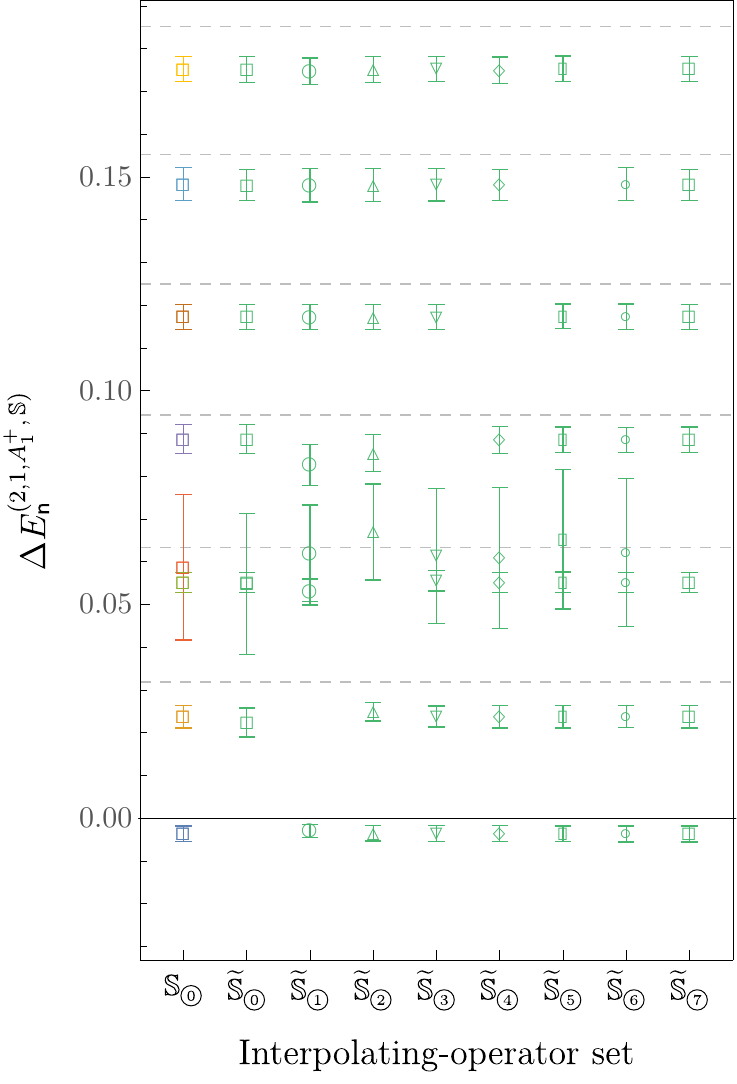}
   \caption{\label{fig:B2I1A1_rainbow} The left panel shows effective FV energy shifts obtained using $\mathbb{S}_{\protect \circled{0}}^{(2,1,A_1^+)}$. Light (dark) shaded regions are $67\%$ bootstrap confidence intervals for final fit results (the highest weight fits), and dashed lines are non-interacting energies.  These results for fitted FV energy shifts are also shown in the leftmost column of the right panel and compared to results obtained using interpolating operator sets $\widetilde{\mathbb{S}}_{\protect \circled{\text{m}}}^{(2,1,A_1^+)}$ in which the operators that have largest overlap with a particular energy level are omitted as discussed in the main text. }
\end{figure}

For $B=2$ systems, correlation-function matrices for $I \in \{0, 1\}$ channels include thin- and wide-smeared hexaquark operators, all linearly independent dibaryon operators with $\mathfrak{s} \leq 6$, and quasi-local operators with exponential localization scales $\kappa_{\mathfrak{q}} \in \{\kappa_1,\kappa_2,\kappa_3\} = \{0.035,0.070,0.14 \}$.
These scales are associated with binding energies of $\{0.0010,0.0041,0.016\}$ in lattice units, corresponding to $\{ 1.4,\ 5.5,\ 22\}$ MeV, which ranges from less than the binding energy of the deuteron in nature to the average of the more deeply-bound results for negative FV energy shifts for the  dineutron and deuteron channels obtained in  Refs.~\cite{Beane:2012vq,Berkowitz:2015eaa,Wagman:2017tmp} for the same gauge-field ensemble.

\begin{figure}[!t]
  \centering
	\includegraphics[width=0.5\columnwidth]{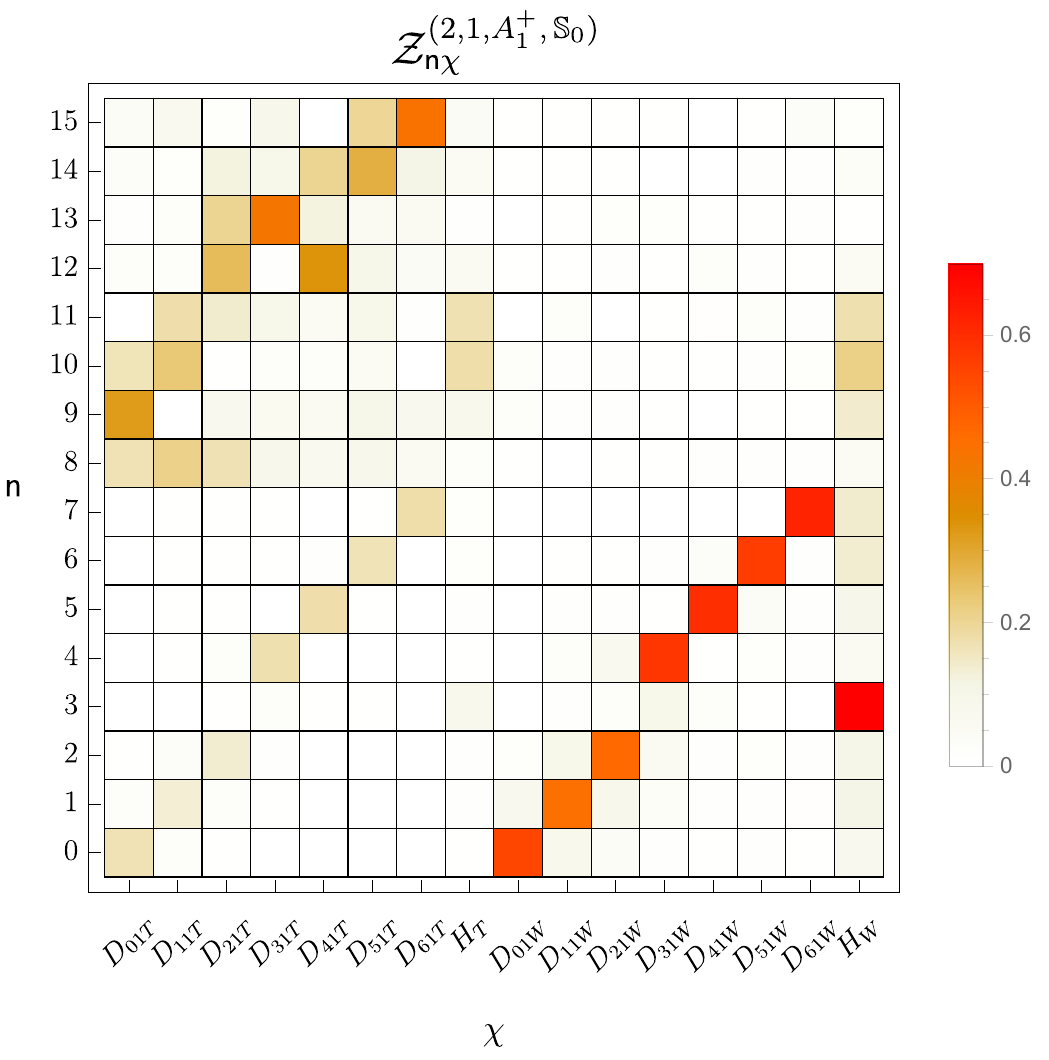}
   \caption{\label{fig:B2I1A1_Zplot} Results for relative overlap factors $\mathcal{Z}_{\mathsf{n} \chi}^{(2,1,A_1^+,\mathbb{S}_0)}$ for dibaryon operators $D_{\mathfrak{s}1g}$ and hexaquark operators $H_g$ with GEVP eigenstates in the dineutron channel.}
\end{figure}

Correlation-function matrices including all 22 interpolating operators with $I=1$ and $\Gamma_J = A_1^+$ are degenerate, that is $\det [C_{\chi\chi'}^{(2,1,A_1^+)}(t)]$ is consistent with zero at $1\sigma$ for all $t$, suggesting that this interpolating-operator set has statistically significant overlap with fewer than 22 LQCD energy eigenstates.
Within the statistical precision of this work, the largest non-degenerate interpolating-operator sets are found to include 16 operators.
One such set, $\mathbb{S}_{\circled{0}}^{(2,1,A_1^+)}$, includes all hexaquark and dibaryon operators with $\mathfrak{s} \leq 6$ but no quasi-local operators.
The FV energy shifts obtained using this interpolating operator set are shown in Fig.~\ref{fig:B2I1A1_rainbow}, and in particular $\Delta E_{0}^{(2,1,A_1^+,\mathbb{S}_0)} = -0.0037(18)$.
Other interpolating operator sets with $\mathfrak{s} \in \{0,1\}$ dibaryon operators replaced by quasi-local interpolating operators give consistent results with larger statistical uncertainties.
However, applying the same fitting procedure to asymmetric correlation functions with hexaquark sources and $\mathfrak{s}=0$ dibaryon sinks gives a corresponding ground-state FV energy shift of $-0.0091(75)$, while results from analogous correlation functions using a superset of this gauge-field ensemble in Ref.~\cite{Beane:2012vq}, Ref.~\cite{Berkowitz:2015eaa}, and Ref.~\cite{Wagman:2017tmp} give results of $-0.0111(21)$, $-0.0127(21)$, and $-0.0137(17)$, respectively.

Subsets of $\mathbb{S}_{\circled{0}}^{(2,1,A_1^+)}$ that in which dibaryon operators with $\mathfrak{s} = \text{m}$ are omitted, denoted $\widetilde{\mathbb{S}}_{\circled{\text{m}}}^{(2,1,A_1^+)}$ with $0 \leq \text{m} \leq 6$, and a set in which hexaquark operators are omitted, denoted $\widetilde{\mathbb{S}}_{\circled{7}}^{(2,1,A_1^+)}$, give energy level determinations that differ from those obtained using $\mathbb{S}_{\circled{0}}^{(2,1,A_1^+)}$ at high statistical significance as shown in Fig.~\ref{fig:B2I1A1_rainbow}.
The energy levels that dominantly overlap with the omitted interpolating operators appear to be missing in results obtained using $\widetilde{\mathbb{S}}_{\circled{\text{m}}}^{(2,1,A_1^+)}$, while the results for other energy levels are unaffected within statistical uncertainties.
Overlap-factor calculations further show that the interpolating-operators with the largest overlap with each state are approximately orthogonal, see Fig.~\ref{fig:B2I1A1_Zplot}.
$\widetilde{\mathbb{S}}_{\circled{0}}^{(2,1,A_1^+)}$ provides a ground-state energy determination that is consistent with the first-excited-state energy determined using  $\mathbb{S}_{\circled{0}}^{(2,1,A_1^+)}$, demonstrating analogous behavior to the model discussed in Sec.~\ref{sec:toy}.
Despite these discrepancies, it is important to note that all results obtained using variational methods are consistent if interpreted as upper bounds on energy levels.

Results for the deuteron channel are analogous. Correlation-function matrices involving all 48 operators with $I=0$ and $\Gamma_J = T_1^+$ considered are found to be degenerate. The largest non-degenerate sets include 42 interpolating operators, for example $\mathbb{S}_{\circled{0}}^{(2,0,T_1^+)}$, which includes all hexaquark and dibaryon but no quasi-local interpolating operators.
Consistent results are obtained when $\mathfrak{s} \in \{0,1\}$ dibaryon operators are replaced with quasi-local operators, but removing operators from $\mathbb{S}_{\circled{0}}^{(2,0,T_1^+)}$ again leads to missing energy levels and a set of 40 interpolating operators misses the ground state identified by $\mathbb{S}_{\circled{0}}^{(2,0,T_1^+)}$. Overlap quality, rather than interpolating operator quantity, is essential to ensure that energy levels are accurately determined using variational methods.

To robustly determine the two-nucleon energy spectrum, future studies must include a wide range of interpolating operators that span as much physically motivated Hilbert space as possible.
Although GEVP results can reconstruct hexaquark-dibaryon correlation functions as linear combinations of states that are closer to threshold than the apparent plateau observed in these correlation functions, it is also possible that these asymmetric correlation functions could reveal an actual state that has smaller overlap than other states with all interpolating operators considered here. % in this work. %, as discussed in Sec.~\ref{sec:toy}.
Further studies of interpolating-operator dependence are needed to assess the systematic uncertainties associated with LQCD calculations of multi-baryon systems.

\acknowledgments{ I gratefully thank my collaborators on the research presented here: Saman Amarasinghe, Riyadh Baghdadi, Zohreh Davoudi, William Detmold, Marc Illa, Assumpta Parre\~no, Andrew Pochinsky, and Phiala Shanahan. I further thank Silas Beane, Kostas Orginos, and Martin Savage for extensive discussions and Martin Savage for collaboration in the early stages of this work.
This research used resources of the Oak Ridge Leadership Computing Facility at the Oak Ridge National Laboratory, which is supported by the Office of Science of the U.S. Department of Energy under Contract number DE-AC05-00OR22725, and the University of Washington Hyak computational infrastructure.
Computations used the \verb!Qlua!~\cite{qlua} and \verb!Tiramisu!~\cite{baghdadi2020tiramisu} software libraries.
This manuscript has been authored by Fermi Research Alliance, LLC under Contract No. DE-AC02-07CH11359 with the U.S. Department of Energy, Office of Science, Office of High Energy Physics.}

\bibliography{variational}

\begin{thebibliography}{10}

\bibitem{Parisi:1983ae}
G.~Parisi.
\newblock {The Strategy for Computing the Hadronic Mass Spectrum}.
\newblock {\em Phys. Rept.}, 103:203--211, 1984.

\bibitem{Lepage:1989hd}
G.~Peter Lepage.
\newblock {The Analysis of Algorithms for Lattice Field Theory}.
\newblock In {\em {Theoretical Advanced Study Institute in Elementary Particle
  Physics}}, 6 1989.

\bibitem{Beane:2009gs}
Silas~R. Beane, William Detmold, Thomas~C Luu, Kostas Orginos, Assumpta
  Parre\~no, Martin~J. Savage, Aaron Torok, and Andre Walker-Loud.
\newblock {High Statistics Analysis using Anisotropic Clover Lattices. II.
  Three-Baryon Systems}.
\newblock {\em Phys. Rev. D}, 80:074501, 2009.

\bibitem{Wagman:2016bam}
Michael~L. Wagman and Martin~J. Savage.
\newblock {Statistics of baryon correlation functions in lattice QCD}.
\newblock {\em Phys. Rev. D}, 96(11):114508, 2017.

\bibitem{Davoudi:2020ngi}
Zohreh Davoudi, William Detmold, Kostas Orginos, Assumpta Parre\~no, Martin~J.
  Savage, Phiala Shanahan, and Michael~L. Wagman.
\newblock {Nuclear matrix elements from lattice QCD for electroweak and
  beyond-Standard-Model processes}.
\newblock {\em Phys. Rept.}, 900:1--74, 2021.

\bibitem{Luscher:1986pf}
Martin L{\"u}scher.
\newblock {Volume Dependence of the Energy Spectrum in Massive Quantum Field
  Theories. 2. Scattering States}.
\newblock {\em Commun. Math. Phys.}, 105:153--188, 1986.

\bibitem{Ishii:2006ec}
N.~Ishii, S.~Aoki, and T.~Hatsuda.
\newblock {The Nuclear Force from Lattice QCD}.
\newblock {\em Phys. Rev. Lett.}, 99:022001, 2007.

\bibitem{Inoue:2010es}
Takashi Inoue, Noriyoshi Ishii, Sinya Aoki, Takumi Doi, Tetsuo Hatsuda, Yoichi
  Ikeda, Keiko Murano, Hidekatsu Nemura, and Kenji Sasaki.
\newblock {Bound H-dibaryon in Flavor SU(3) Limit of Lattice QCD}.
\newblock {\em Phys. Rev. Lett.}, 106:162002, 2011.

\bibitem{Aoki:2020bew}
Sinya Aoki and Takumi Doi.
\newblock {Lattice QCD and baryon-baryon interactions: HAL QCD method}.
\newblock {\em Front. in Phys.}, 8:307, 2020.

\bibitem{Birse:2012ph}
Michael~C. Birse.
\newblock {Potential problems with interpolating fields}.
\newblock {\em Eur. Phys. J. A}, 53(11):223, 2017.

\bibitem{Beane:2010em}
S.~R. Beane, W.~Detmold, K.~Orginos, and M.~J. Savage.
\newblock {Nuclear Physics from Lattice QCD}.
\newblock {\em Prog. Part. Nucl. Phys.}, 66:1--40, 2011.

\bibitem{Yamazaki:2017gjl}
Takeshi Yamazaki and Yoshinobu Kuramashi.
\newblock {Relation between scattering amplitude and Bethe-Salpeter wave
  function in quantum field theory}.
\newblock {\em Phys. Rev. D}, 96(11):114511, 2017.

\bibitem{Iritani:2018zbt}
Takumi Iritani, Sinya Aoki, Takumi Doi, Shinya Gongyo, Tetsuo Hatsuda, Yoichi
  Ikeda, Takashi Inoue, Noriyoshi Ishii, Hidekatsu Nemura, and Kenji Sasaki.
\newblock {Systematics of the HAL QCD Potential at Low Energies in Lattice
  QCD}.
\newblock {\em Phys. Rev. D}, 99(1):014514, 2019.

\bibitem{Drischler:2019xuo}
Christian Drischler, Wick Haxton, Kenneth McElvain, Emanuele Mereghetti, Amy
  Nicholson, Pavlos Vranas, and Andr\'e Walker-Loud.
\newblock {Towards grounding nuclear physics in QCD}.
\newblock 10 2019.

\bibitem{Fox:1981xz}
G.~Fox, R.~Gupta, O.~Martin, and S.~Otto.
\newblock {Monte Carlo Estimates of the Mass Gap of the O(2) and O(3) Spin
  Models in (1+1)-dimensions}.
\newblock {\em Nucl. Phys. B}, 205:188--220, 1982.

\bibitem{Michael:1982gb}
Christopher Michael and I.~Teasdale.
\newblock {Extracting Glueball Masses From Lattice QCD}.
\newblock {\em Nucl. Phys. B}, 215:433--446, 1983.

\bibitem{Luscher:1990ck}
Martin L{\"u}scher and Ulli Wolff.
\newblock {How to Calculate the Elastic Scattering Matrix in Two-dimensional
  Quantum Field Theories by Numerical Simulation}.
\newblock {\em Nucl. Phys. B}, 339:222--252, 1990.

\bibitem{Allton:1993wc}
C.~R. Allton et~al.
\newblock {Gauge invariant smearing and matrix correlators using Wilson
  fermions at Beta = 6.2}.
\newblock {\em Phys. Rev. D}, 47:5128--5137, 1993.

\bibitem{Burch:2004he}
Tommy Burch, Christof Gattringer, Leonid~Ya. Glozman, Reinhard Kleindl, C.~B.
  Lang, and Andreas Schaefer.
\newblock {Spatially improved operators for excited hadrons on the lattice}.
\newblock {\em Phys. Rev. D}, 70:054502, 2004.

\bibitem{Peardon:2009gh}
Michael Peardon, John Bulava, Justin Foley, Colin Morningstar, Jozef Dudek,
  Robert~G. Edwards, Balint Jo\'o, Huey-Wen Lin, David~G. Richards, and
  Keisuke~Jimmy Juge.
\newblock {A Novel quark-field creation operator construction for hadronic
  physics in lattice QCD}.
\newblock {\em Phys. Rev. D}, 80:054506, 2009.

\bibitem{Morningstar:2011ka}
Colin Morningstar, John Bulava, Justin Foley, Keisuke~J. Juge, David Lenkner,
  Mike Peardon, and Chik~Him Wong.
\newblock {Improved stochastic estimation of quark propagation with Laplacian
  Heaviside smearing in lattice QCD}.
\newblock {\em Phys. Rev. D}, 83:114505, 2011.

\bibitem{Dudek:2012xn}
Jozef~J. Dudek, Robert~G. Edwards, and Christopher~E. Thomas.
\newblock {Energy dependence of the $\rho$ resonance in $\pi\pi$ elastic
  scattering from lattice QCD}.
\newblock {\em Phys. Rev. D}, 87(3):034505, 2013.
\newblock [Erratum: Phys.Rev.D 90, 099902 (2014)].

\bibitem{Wilson:2015dqa}
David~J. Wilson, Raul~A. Brice\~{n}o, Jozef~J. Dudek, Robert~G. Edwards, and
  Christopher~E. Thomas.
\newblock {Coupled $\pi\pi, K\bar{K}$ scattering in $P$-wave and the $\rho$
  resonance from lattice QCD}.
\newblock {\em Phys. Rev. D}, 92(9):094502, 2015.

\bibitem{Lang:2012db}
C.~B. Lang and V.~Verduci.
\newblock {Scattering in the \ensuremath{\pi}N negative parity channel in
  lattice QCD}.
\newblock {\em Phys. Rev. D}, 87(5):054502, 2013.

\bibitem{Kiratidis:2015vpa}
Adrian~L. Kiratidis, Waseem Kamleh, Derek~B. Leinweber, and Benjamin~J. Owen.
\newblock {Lattice baryon spectroscopy with multi-particle interpolators}.
\newblock {\em Phys. Rev. D}, 91:094509, 2015.

\bibitem{Beane:2010hg}
S.~R. Beane et~al.
\newblock {Evidence for a Bound H-dibaryon from Lattice QCD}.
\newblock {\em Phys. Rev. Lett.}, 106:162001, 2011.

\bibitem{Beane:2011iw}
S.~R. Beane, E.~Chang, W.~Detmold, H.~W. Lin, T.~C. Luu, K.~Orginos,
  A.~Parre\~no, M.~J. Savage, A.~Torok, and A.~Walker-Loud.
\newblock {The Deuteron and Exotic Two-Body Bound States from Lattice QCD}.
\newblock {\em Phys. Rev. D}, 85:054511, 2012.

\bibitem{Beane:2012vq}
S.~R. Beane, E.~Chang, S.~D. Cohen, William Detmold, H.~W. Lin, T.~C. Luu,
  K.~Orginos, A.~Parre\~{n}o, M.~J. Savage, and A.~Walker-Loud.
\newblock {Light Nuclei and Hypernuclei from Quantum Chromodynamics in the
  Limit of SU(3) Flavor Symmetry}.
\newblock {\em Phys. Rev. D}, 87(3):034506, 2013.

\bibitem{Yamazaki:2012hi}
Takeshi Yamazaki, Ken-ichi Ishikawa, Yoshinobu Kuramashi, and Akira Ukawa.
\newblock {Helium nuclei, deuteron and dineutron in 2+1 flavor lattice QCD}.
\newblock {\em Phys. Rev. D}, 86:074514, 2012.

\bibitem{Beane:2013br}
S.~R. Beane, E.~Chang, S.~D. Cohen, W.~Detmold, P.~Junnarkar, H.~W. Lin, T.~C.
  Luu, K.~Orginos, A.~Parre\~no, M.~J. Savage, and A.~Walker-Loud.
\newblock {Nucleon-Nucleon Scattering Parameters in the Limit of SU(3) Flavor
  Symmetry}.
\newblock {\em Phys. Rev. C}, 88(2):024003, 2013.

\bibitem{Berkowitz:2015eaa}
Evan Berkowitz, Thorsten Kurth, Amy Nicholson, Balint Jo\'o, Enrico Rinaldi,
  Mark Strother, Pavlos~M. Vranas, and Andr\'e Walker-Loud.
\newblock {Two-Nucleon Higher Partial-Wave Scattering from Lattice QCD}.
\newblock {\em Phys. Lett. B}, 765:285, 2017.

\bibitem{Yamazaki:2015asa}
Takeshi Yamazaki, Ken-ichi Ishikawa, Yoshinobu Kuramashi, and Akira Ukawa.
\newblock {Study of quark mass dependence of binding energy for light nuclei in
  2+1 flavor lattice QCD}.
\newblock {\em Phys. Rev. D}, 92(1):014501, 2015.

\bibitem{Wagman:2017tmp}
Michael~L. Wagman, Frank Winter, Emmanuel Chang, Zohreh Davoudi, William
  Detmold, Kostas Orginos, Martin~J. Savage, and Phiala~E. Shanahan.
\newblock {Baryon-Baryon Interactions and Spin-Flavor Symmetry from Lattice
  Quantum Chromodynamics}.
\newblock {\em Phys. Rev. D}, 96(11):114510, 2017.

\bibitem{Doi:2012xd}
Takumi Doi and Michael~G. Endres.
\newblock {Unified contraction algorithm for multi-baryon correlators on the
  lattice}.
\newblock {\em Comput. Phys. Commun.}, 184:117, 2013.

\bibitem{Detmold:2012eu}
William Detmold and Kostas Orginos.
\newblock {Nuclear correlation functions in lattice QCD}.
\newblock {\em Phys. Rev. D}, 87(11):114512, 2013.

\bibitem{Francis:2018qch}
A.~Francis, J.R. Green, P.M. Junnarkar, Ch. Miao, T.D. Rae, and H.~Wittig.
\newblock {Lattice QCD study of the $H$ dibaryon using hexaquark and two-baryon
  interpolators}.
\newblock {\em Phys. Rev. D}, 99(7):074505, 2019.

\bibitem{Horz:2020zvv}
Ben H\"orz et~al.
\newblock {Two-nucleon S-wave interactions at the $SU(3)$ flavor-symmetric
  point with $m_{ud}\simeq m_s^{\rm phys}$: A first lattice QCD calculation
  with the stochastic Laplacian Heaviside method}.
\newblock {\em Phys. Rev. C}, 103(1):014003, 2021.

\bibitem{Green:2021qol}
Jeremy~R. Green, Andrew~D. Hanlon, Parikshit~M. Junnarkar, and Hartmut Wittig.
\newblock {Weakly bound $H$ dibaryon from SU(3)-flavor-symmetric QCD}, 3 2021.

\bibitem{Detmold:2019fbk}
W.~Detmold, D.~J. Murphy, A.~V. Pochinsky, M.~J. Savage, P.~E. Shanahan, and
  M.~L. Wagman.
\newblock {Sparsening algorithm for multihadron lattice QCD correlation
  functions}.
\newblock {\em Phys. Rev. D}, 104(3):034502, 2021.

\bibitem{Li:2020hbj}
Yuan Li, Shi-Cheng Xia, Xu~Feng, Lu-Chang Jin, and Chuan Liu.
\newblock {Field sparsening for the construction of the correlation functions
  in lattice QCD}.
\newblock {\em Phys. Rev. D}, 103(1):014514, 2021.

\bibitem{baghdadi2020tiramisu}
Riyadh Baghdadi, Abdelkader~Nadir Debbagh, Kamel Abdous, Fatima~Zohra
  Benhamida, Alex Renda, Jonathan~Elliott Frankle, Michael Carbin, and Saman
  Amarasinghe.
\newblock Tiramisu: A polyhedral compiler for dense and sparse deep learning,
  2020.

\bibitem{Amarasinghe:2021lqa}
Saman Amarasinghe, Riyadh Baghdadi, Zohreh Davoudi, William Detmold, Marc Illa,
  Assumpta Parre\~no, Andrew~V. Pochinsky, Phiala~E. Shanahan, and Michael~L.
  Wagman.
\newblock {A variational study of two-nucleon systems with lattice QCD}.
\newblock 8 2021.

\bibitem{Basak:2005ir}
Subhasish Basak, Robert Edwards, George~T. Fleming, Urs~M. Heller, Colin
  Morningstar, David Richards, Ikuro Sato, and Stephen~J. Wallace.
\newblock {Clebsch-Gordan construction of lattice interpolating fields for
  excited baryons}.
\newblock {\em Phys. Rev. D}, 72:074501, 2005.

\bibitem{Gusken:1989ad}
S.~G{\"u}sken, U.~Low, K.H. Mutter, R.~Sommer, A.~Patel, and K.~Schilling.
\newblock {Nonsinglet Axial Vector Couplings of the Baryon Octet in Lattice
  {QCD}}.
\newblock {\em Phys. Lett. B}, 227:266--269, 1989.

\bibitem{Luu:2011ep}
Thomas Luu and Martin~J. Savage.
\newblock {Extracting Scattering Phase-Shifts in Higher Partial-Waves from
  Lattice QCD Calculations}.
\newblock {\em Phys. Rev. D}, 83:114508, 2011.

\bibitem{Bulava:2016mks}
John Bulava, Brendan Fahy, Ben H\"orz, Keisuke~J. Juge, Colin Morningstar, and
  Chik~Him Wong.
\newblock {$I=1$ and $I=2$ $\pi-\pi$ scattering phase shifts from
  $N_{\mathrm{f}} = 2+1$ lattice QCD}.
\newblock {\em Nucl. Phys. B}, 910:842--867, 2016.

\bibitem{Lepage:1992xa}
G.~Peter Lepage and Paul~B. Mackenzie.
\newblock {On the viability of lattice perturbation theory}.
\newblock {\em Phys. Rev. D}, 48:2250--2264, 1993.

\bibitem{Luscher:1984xn}
{M. L\"{u}scher and P. Weisz}.
\newblock {On-Shell Improved Lattice Gauge Theories}.
\newblock {\em Commun. Math. Phys.}, 97:59, 1985.
\newblock [Erratum: Commun. Math. Phys.98,433(1985)].

\bibitem{Sheikholeslami:1985ij}
B.~Sheikholeslami and R.~Wohlert.
\newblock {Improved Continuum Limit Lattice Action for QCD with Wilson
  Fermions}.
\newblock {\em Nucl. Phys.}, B259:572, 1985.

\bibitem{Morningstar:2003gk}
Colin Morningstar and Mike~J. Peardon.
\newblock {Analytic smearing of SU(3) link variables in lattice QCD}.
\newblock {\em Phys. Rev. D}, 69:054501, 2004.

\bibitem{qlua}
Andrew Pochinsky.
\newblock {Qlua. https://usqcd.lns.mit.edu/qlua}.

\bibitem{Beane:2020ycc}
S.~R. Beane, W.~Detmold, R.~Horsley, M.~Illa, M.~Jafry, D.~J. Murphy,
  Y.~Nakamura, H.~Perlt, P.~E.~L. Rakow, G.~Schierholz, P.~E. Shanahan,
  H.~St{\"u}ben, M.~L. Wagman, F.~Winter, R.~D. Young, and J.~M. Zanotti.
\newblock {Charged multi-hadron systems in lattice QCD+QED}.
\newblock {\em Phys. Rev. D}, 103:054504, 2021.

\end{thebibliography}

\bibliographystyle{unsrt}

%\begin{thebibliography}{99}
%\bibitem{...}
%....
%\end{thebibliography}

\end{document}